\documentclass[11pt,a4paper]{article}
\usepackage{amsmath,amsfonts,amssymb,mathtools,amsthm,newlfont,graphicx,amscd,bbm,enumerate,galois,mathrsfs,xypic,geometry,bbm,xypic}
\usepackage{booktabs,multirow,bigstrut,tabularx,extarrows,simplewick}
\usepackage{tikz}

\usepackage{hyperref}

\usepackage{color}

\newcommand*{\fps}[1]{[\![#1]\!]}        

\newcommand{\lmd}{\lambda}

\newcommand{\Lmd}{\Lambda}

\newcommand{\p}{\partial}

\newcommand{\veps}{\varepsilon}
\newcommand{\fai}{\varphi}

\newcommand{\rme}{{\mathrm e}}

\newcommand{\mfkc}{\mathfrak{c}}

\newcommand{\mcalA}{\mathcal{A}}

\newcommand{\mcalF}{\mathcal{F}}

\newcommand{\mcalJ}{\mathcal{J}}
\newcommand{\mcalK}{\mathcal{K}}

\newcommand{\mcalP}{\mathcal{P}}

\newcommand{\bbC}{\mathbb{C}}

\newcommand{\bbZ}{\mathbb{Z}}

\newcommand{\bsv}{\boldsymbol{v}}

\newcommand{\bfP}{\mathbf{P}}
\newcommand{\sfA}{\mathsf{A}}

\newcommand{\bsu}{\boldsymbol{u}}

\newcommand*{\pp}[1]
  {\frac{\partial   }
        {\partial #1}
  }

\newcommand*{\pfrac}[2]
  {\frac{\partial #1}
        {\partial #2}
  }

\DeclareMathOperator{\End}{End}

\DeclareMathOperator*{\Res}{Res}       

\DeclareMathOperator{\td}{d\!}

\newtheorem{thm}{Theorem}[section]

\newtheorem{rmk}[thm]{Remark}

\newtheorem{prop}[thm]{Proposition}

\newtheorem{ex}[thm]{Example}

\numberwithin{equation}{section}
\allowdisplaybreaks[4]

\begin{document}

\title{{Central invariants for certain lattice hierarchies via generating kernels of bihamiltonian structures  
}
}
	\author{{
		Kai Liu,~ Haonan Qu,~ Chao-Zhong Wu} \\[1em]
		\footnotesize 
         School of Mathematics, Sun Yat-Sen University,
		 Guangzhou 510275, China \\[10pt]
\footnotesize
Emails: liuk233@mail2.sysu.edu.cn,\, quhn@mail.sysu.edu.cn,\, wuchaozhong@sysu.edu.cn
	}
    \date{}
	\maketitle
	\begin{abstract}
For any positive integer $n$, we prove that the central invariants of the bihamiltonian structure for the
$(n,1)$-type rational reduction of the 2D-Toda hierarchy are equal to $\frac{1}{24}$.
Our method is based on certain generating kernels of bihamiltonian structures, which admit
closed-form expressions in terms of the symbol of Lax operator,
and encode the deformation data needed to compute the central invariants.
This method is also applied to other lattice hierarchies, and shows that the central invariants of the
bihamiltonian structures for the extended bi-graded Toda hierarchy and for the
$q$-deformed Gelfand--Dickey hierarchy are equal to $\frac{1}{24}$, which agree with the results in literature via a different approach.
	\end{abstract}
{\footnotesize
\\[6pt]
}


\section{Introduction}

For a semisimple bihamiltonian structure of hydrodynamic type, Dubrovin, Liu and Zhang \cite{Hamiltonian perturbation,LiuZhang2005JGP} showed that all dispersive deformations are classified up to Miura-type transformations by a collection of single-variable functions $\mfkc_i(u^i)$, 
called \textit{central invariants}, of certain canonical coordinates $u^i$.  
These invariants play a critical role in the theory of topological deformations of the principal hierarchies associated to semisimple Dubrovin--Frobenius manifolds \cite{2DTFT}. 
These deformed hierarchies, named as the Dubrovin--Zhang hierarchies \cite{normal form} 
are equipped with bihamiltonian structures whose central invariants $\mfkc_i(u^i)=\frac{1}{24}$. 
It hints that central invariants may provide a criterion 
to find candidates of bihamiltonian integrable hierarchies related to topological field theory.

Explicit central invariants have been calculated for several important classes of integrable hierarchies,
such as the Drinfeld--Sokolov hierarchies \cite{CentralInv2008adv},
the bi-graded Toda hierarchy \cite{CDZcom},
the constrained KP hierarchy \cite{central inv of cKP}
and the $q$-deformed Gelfand--Dickey hierarchy \cite{rational-reduction, Frenkel-qKdV, Romano}, etc.
The method is technical in a sense,
which involves certain multiple integral representations of the bihamiltonian structures and star product between symbols of Lax operators.

In the recent decade, the $(n,m)$-type rational reduction of the 2D Toda hierarchy ($(n,m)$-RR2T), with two positive integers $n$ and $m$, has attracted much interest of researchers. This is a lattice hierarchy defined by a Lax operator of the form
\[
L=(1+W_1\Lambda^{-1}+\dots+W_m\Lambda^{-m})^{-1}(\Lambda^n+U_1\Lambda^{n-1}+\dots+U_{n-1}\Lambda+U_n),
\]
where $U_1,\dots,U_n, W_1,\dots,W_m$ are unknown functions depending on some spatial variable $x$, and  $\Lambda=\rme^{\veps \partial_x}$ is a shift operator. As shown by Brini, Carlet, Romano and Rossi \cite{rational-reduction}, the $(n,m)$-RR2T  carries a Hamiltonian structure degenerated by an extended version of the ``second'' Hamilton structure for the 2D Toda hierarchy \cite{trh-toda} (cf. \cite{Rmat-Wu}), and its tau function can serve as a descendent Gromov--Witten potential up to genus one. In case $n=m=1$, this hierarchy, also known as the Ablowitz--Ladik hierarchy, is found to be tri-Hamiltonian \cite{Kupershmidt,AL-triham,OFZR}; when $n\ge2$, the bihamiltonian structure for the $(n,1)$-RR2T was constructed in \cite{ZQL-n1RR2T}. Moreover, the central invariants for the $(n,1)$-RR2T with $n=1,2$ have been computed in \cite{AL-triham} and \cite{2-1 RR2T} respectively.

In this paper, we want to calculate the central invariants for the $(n,1)$-RR2T with arbitrary $n\ge1$, and achieve the following main result (see Theorem \ref{thm:main thm} below).
\begin{thm} \label{mainthm}
For any positive integer $n$, the central invariants of the bihamiltonian structure \cite{ZQL-n1RR2T} for the
$(n,1)$-RR2T are equal to $\frac{1}{24}$.
\end{thm}
\noindent
In particular, when $n=1,2$, this result coincides with that obtained in \cite{AL-triham,2-1 RR2T}.
Our approach is different from those used in the literature. More exactly, we introduce the so-called generating kernels (see \eqref{G-ell-pqe} below) of the bihamiltonian structure, which admit closed-form expressions in terms of symbol of the Lax operator $L$. These  generating kernels encode the deformation data needed to evaluate the central invariants in canonical coordinates.

Our motivation is also from the relationship between integrable hierarchies and generalized Frobenius manifolds with non-flat unit vector field \cite{Liu2024,GFM}. In \cite{extended AL}, Liu, Wang and Zhang proved that the  Ablowitz--Ladik hierarchy ($(1,1)$-RR2T) can be extended to the topological deformation of the Principal Hierarchy associated to a generalized Frobenius manifold (cf. \cite{Brini-AL}). Along this line, for $n\ge2$, Qu and Zhao \cite{ZQL-n1RR2T} showed that the dispersionless flows of the $(n,1)$-RR2T coincide with a subfamily of the flows of the Principal Hierarchy associated to a generalized Frobenius manifold $\mathcal{M}_{(n,1)}$. Under a polynomiality conjecture for generalized Frobenius manifolds, the deformed bihamiltonian structure is expected to have all central invariants equal to $\frac{1}{24}$. Our computation for the $(n,1)$-RR2T provides evidence that the $(n,1)$-RR2T may be a subhierarchy of the corresponding topological deformation of the Principal Hierarchy  associated to $\mathcal{M}_{(n,1)}$ for $n\ge1$.

Moreover, the generating-kernel method developed in the present paper
is also applied to the extended bi-graded Toda hierarchy and the $q$-deformed Gelfand--Dickey hierarchy.
It is effective to show that their central invariants are also equal to $\frac{1}{24}$,
which are consistent with the known values obtained in \cite{CDZcom,CZL} via a different approach.
Finally, we also give an explicit Miura-type transformation
relating the bihamiltonian structure for the $(n,1)$-RR2T to a modification of that for the $(n,1)$-type extended bi-graded Toda hierarchy.

The paper is organized as follows.
Section 2 recalls the definition of central invariants of a bihamiltonian structure. In Section 3, we recall the bihamiltonian structure for the $(n,1)$-type RR2T, and prove Theorem \ref{mainthm} via the generating kernels.
In Section 4, we apply the generating kernel  method to the extended bi-graded Toda and $q$-deformed Gelfand--Dickey hierarchies.
Section 5 is devoted to conclusions and remarks.
In the appendix, we study the relationship between the bihamiltonian structure for the $(n,1)$-RR2T and that for the $(n,1)$-type extended bi-graded Toda hierarchy.

\section{Definition of central invariants of a bihamiltonian structure}
\label{subsection:def of central inv}

Let $M$ be an $m$-dimensional manifold with local coordinates $\bsv=(v^1,\dots, v^m)$.
Suppose that $\bsv$ depends smoothly on a spatial variable $x$.
Let $\mcalA$ be the ring of complete differential polynomials generated by $\bsv$.
The space of local functionals is denoted by $\mcalF:=\mcalA/\p_x\mcalA$, whose element $f+\p_x\mcalA$ is written as $\int f \td x$.
We assume that the following two brackets
\begin{align}
  \{,\}_\ell\colon \mcalF\times\mcalF \to\, \mcalF, \quad
  (F,G) \mapsto\,
  \int \sum_{i,j=1}^{m} \frac{\delta F}{\delta v^i}
  \left(\mcalP_\ell^{ij}
  \frac{\delta G}{\delta v^j}\right)\td x,\quad \ell=1,2,
\end{align}
form a pair of compatible Poisson brackets, i.e. a bihamiltonian structure.
%
Here the operators $\mcalP_\ell^{ij}$ take the form
\begin{equation}\label{general HamP}
  \mcalP^{ij}_{\ell}=\sum_{k\geq 0}\veps^k
    \sum_{s=0}^{k+1}
      \sfA_{k,s;\ell}^{ij}(\bsv,\bsv_x,\dots,\p_x^s\bsv)\p_x^{k-s+1}, \quad \ell=1,2,
\end{equation}
with coefficients $\sfA_{k,s;\ell}^{ij}\in \mcalA$ being
homogeneous of differential degree $s$. In particular, for each $k\geq 0$ and $\ell=1,2$, the functions $\sfA_{k,0;\ell}^{ij}$
define a $(2,0)$-tensor field on $M$ as
\begin{equation}\label{tensor-Akl}
  \sfA_{k;\ell} = \sum_{i,j=1}^{m}
  \sfA_{k,0;\ell}^{ij}\pp{v^i}\otimes\pp{v^j}.
\end{equation}
Among them the leading fields
\begin{equation}\label{gl=A0l}
g_\ell:=\sfA_{0;\ell}
\end{equation}
are the contravariant metrics associated to $\mcalP_\ell$.

The bihamiltonian structure $(\mcalP_1,\mcalP_2)$
is called semisimple if the roots $u^1(\bsv),\dots,u^m(\bsv)$ of the
characteristic polynomial
\[
\lmd\mapsto\det\left(g_2^{ij}-\lmd g_1^{ij}\right)
\]
form a system of local coordinates around a generic point of $M$.
These local coordinates $\bsu:=(u^1,\dots u^m)$ are called the \textit{canonical coordinates} of the bihamiltonian structure.
In the canonical coordinates,  the metrics $g_\ell$ have the form
\begin{equation}\label{diag fi}
  g_1(\td u^i, \td u^j) = \delta_{ij} f^i,\quad
  g_2(\td u^i, \td u^j) = \delta_{ij} u^i f^i
\end{equation}
for certain functions $\{f^i\}_{1\leq i\leq m}$.

Denote
\begin{equation}\label{Hij and Kij}
  H^{ij}_\ell := \sfA_{1;\ell}\left(\td u^i, \td u^j\right),\quad
  K^{ij}_\ell := \sfA_{2;\ell}\left(\td u^i, \td u^j\right), \quad \ell=1,2,
\end{equation}
then the set $\{\mfkc_i\}_{i=1}^m$ of \textit{central invariants}
of the bihamiltonian structure $(\mcalP_1,\mcalP_2)$ is defined by \cite{Hamiltonian perturbation,LiuZhang2005JGP}
\begin{equation}\label{def of cInv}
  \mfkc_i=
  \frac{1}{3(f^i)^2}
  \left(
    K_2^{ii}-u^iK_1^{ii}
   +\sum_{k\neq i}
     \frac{(H_2^{ki}-u^iH_1^{ki})^2}{(u^k-u^i)f^k}
  \right).
\end{equation}
Each $\mfkc_i$ turns out to be a single-variable function depending only on $u^i$, i.e. $\mfkc_i=\mfkc_i(u^i)$.
It is known that two deformations of a hydrodynamic-type bihamiltonian structure are Miura-equivalent if and only if they have the same central invariants.

\section{Central invariants of bihamiltonian structure of $(n,1)$-RR2T}

Given an arbitrary positive integer $n$, a bihamiltonian structure of the $(n,1)$-type rational reduction of the 2D Toda hierarchy was constructed in \cite{ZQL-n1RR2T} (see also \cite{AL-triham, 2-1 RR2T} for the cases $n=1,2$). We are to calculate the central invariants of these bihamiltonian structures via a generating kernel approach.

\subsection{Bihamiltonian structure of $(n,1)$-RR2T}

Let us recall some notations introduced in Sect.\,2 of \cite{ZQL-n1RR2T}.

Given an integer $n \geq 1$, we let
\begin{equation}
  \bsv:=(v^1,\dots,v^{n+1}) = (U_1, \dots, U_n ; W)
\end{equation}
be a family of local coordinates on an $(n+1)$-dimensional manifold.
As above, one has the the ring  $\mcalA$ of complete differential polynomials generated by $\bsv$, and the space  $\mcalF:=\mcalA/\p_x\mcalA$ of local functionals.

Introduce a shift operator $\Lmd:=\rme^{\veps\p_x}\in\End(\mcalA)$ with some parameter $\veps$.
For each $f\in\mcalA$ and $s\in\bbZ$, we write
\[f^{[s]}:=\Lmd^s(f):=\sum_{k=0}^{\infty}\frac{(\veps s \p_x)^k}{k!}f.\]
Consider an operator of the form
\begin{equation}\label{n1RR2T-Lax}
L:=\,
  \left(1-W\Lmd^{-1}\right)^{-1}
  \left(
    \Lmd^n + \sum_{k=1}^{n} U_{k}\Lmd^{n-k}
  \right)\in\mcalA(\!(\Lmd^{-1})\!).
\end{equation}
The $(n,1)$-type \textit{rational reduction of the 2D-Toda hierarchy} ( $(n,1)$-RR2T for short)
\cite{rational-reduction, ZQL-n1RR2T} is defined to be the following system of evolutionary equations:
\begin{align}
  \veps\pfrac{L}{t^{i,k}} &=\,
  \left[
    \left(L^{k+\frac{i}{n}}\right)_+ , L
  \right], \label{RR2T positive flow}
\\
  \veps\pfrac{L}{t^{0,-k-1}} &=\,
    \left[
      \left(L^{-k-1}\right)_- , L
    \right],
\label{RR2T negative flow}
\end{align}
for $1\leq i\leq n$ and $k\geq 0$.
More details, such as
the definitions of $L^{\frac{1}{n}}\in\mcalA(\!(\Lmd^{-1})\!)$ and $L^{-1}\in\mcalA(\!(\Lmd)\!)$,
can be found in \cite{ZQL-n1RR2T} and are omitted here.

For convenience, we introduce the following additional notations:
\begin{equation}\label{short notation}
    U_0 := 1, \quad
    U_k := 0 \quad \text{if\, $k<0$ or $k> n$}.
\end{equation}
In particular, we emphasize that $U_{n+1} = 0$, which is NOT equal to $W$. Such notations will be used throughout the current section.

As was shown in \cite{ZQL-n1RR2T}, the $(n,1)$-RR2T hierarchy \eqref{RR2T positive flow}--\eqref{RR2T negative flow} can be represented in a bihamiltonian form as
\begin{equation}
    \veps\pp{t^{i,k}}
    \begin{pmatrix}
      U_1 \\
      \vdots \\
      U_n \\
      W
    \end{pmatrix}
  =
    \mcalP_1
    \begin{pmatrix}
      \delta H_{i,k}/\delta U_1 \\
      \vdots \\
      \delta H_{i,k}/\delta U_n \\
      \delta H_{i,k}/\delta W
    \end{pmatrix}
  =
    \mcalP_2
    \begin{pmatrix}
      \delta H_{i,k-1}/\delta U_1 \\
      \vdots \\
      \delta H_{i,k-1}/\delta U_n \\
      \delta H_{i,k-1}/\delta W
    \end{pmatrix}
  \end{equation}
with $i\in\{1,2,\dots, n\}$, $k\geq 0$ and $i=0$, $k\leq -1$.
Here the two operator-valued matrices
\begin{equation}\label{n1RR2T Pell}
\mcalP_\ell =\left(\mcalP_\ell^{ij}\right)_{1\leq i,j\leq n+1} \in \mcalA\fps{\veps\p_x}\otimes\bbC^{(n+1)\times (n+1)}, \quad \ell = 1,2,
\end{equation}
are defined by
\begin{align}
  \mcalP_1^{ij}&=\,
    \boldsymbol{1}_{i<n}\boldsymbol{1}_{j<n}\Big(
     U_{i+j-n}\Lmd^{n-i} - \Lmd^{j-n}U_{i+j-n}
    \Big)\notag \\
  &\qquad
    +\Big(
      W\Lmd^{j-n-1}U_{i+j-n-1}-U_{i+j-n-1}\Lmd^{n+1-i}W
    \Big),  \label{260720-1846-1}
\\
  \mcalP_1^{i,n+1} &=\, \delta_{i,n}(\Lmd-1)W,
  \quad
  \mcalP_1^{n+1,i} = \delta_{i,n}W(1-\Lmd^{-1}),
  \quad
  \mcalP_1^{n+1,n+1} = 0; \label{HamP1-afa-W}
\\
  \mcalP_2^{ij} &=\,
    U_i\frac{(\Lmd^{n-i}-1)(1-\Lmd^j)}{\Lmd^n-1}U_j
     +\sum_{s=1}^{j}
     \left(
       U_{j-s}\Lmd^sU_{i+s} - U_{i+s}\Lmd^{j-i-s}U_{j-s}
     \right),
  \label{HamP2Op1}
\\
  \mcalP_2^{i,n+1} &=\,
    U_i\frac{(\Lmd^n-\Lmd^{n-i})(\Lmd-1)}{\Lmd^n-1}W,
  \quad
  \mcalP_2^{n+1,i} = W\frac{(\Lmd^i-1)(1-\Lmd^{-1})}{\Lmd^n-1}U_i, \label{HamP2-afa-W}
\\
  \mcalP_2^{n+1,n+1} &=\, W\frac{(\Lmd^n-\Lmd^{-1})(\Lmd-1)}{\Lmd^n-1}W \label{260720-1846-2}
\end{align}
for $1\leq i,j\leq n$, where $\delta_{ij}$ denotes the Kronecker delta, and
\begin{equation}\label{oneS}
\boldsymbol
1_{\mathrm{P}}:=\begin{cases}
1, & \text{if $\mathrm{P}$ is true}; \\
0, & \text{if $\mathrm{P}$ is false}.
\end{cases}
\end{equation}
The hamiltonian functionals $H_{i,k}\in\mcalF$ are given by
\begin{align*}
  H_{i,k} &:=\, \frac{1}{k+1+\frac{i}{n}}\int\Res\left(L^{k+1+\frac{i}{n}}\right)\td x,
  \quad k\geq -1 \hbox{ and } 1\leq i\leq n; \\
  H_{0,-k-1} &:=\,
    \frac{1}{k}\int\Res\left(L^{-k}\right)\td x,\quad k\geq 1; \\
  H_{0,-1} &:=\, \int\log\frac{W}{U_n}\td x,
\end{align*}
where $\Res\sum_{l}a_l \Lambda^l=a_0$.

\begin{rmk}
As was shown by Brini et al. (see Lemma 2.17 in \cite{rational-reduction}), the second Hamiltonian structure $\mcalP_2$  can also be obtained by restriction of a Hamiltonian structure of the 2D Toda hierarchy \cite{trh-toda} (see also \cite{Rmat-Wu}).
\end{rmk}

\subsection{Generating kernel of Hamiltonian structure}
\label{subsection:genKer}

One observes that the bihamiltonian structure $(\mcalP_1, \mcalP_2)$ given by
\eqref{260720-1846-1}--\eqref{260720-1846-2} has the form \eqref{general HamP}
after the rescaling
\begin{equation}\label{rescaling}
\mcalP_\ell\mapsto \frac{1}{\veps}\mcalP_\ell.
\end{equation}
For example,
the following equalities hold true when $0\leq i,j\leq n-1$:
\begin{align}
  \sfA_{0;1}\left(\td U_{n-i}, \td U_{n-j}\right)
&=\,
  (i+j)U_{n-i-j}\boldsymbol{1}_{i>0}\boldsymbol{1}_{j>0}
 -(i+j+2)WU_{n-1-i-j}, \label{A01}
\\
  \sfA_{0;2}\left(\td U_{n-i}, \td U_{n-j}\right)
&=\,
  \left(
    \frac{ij}{n}-i
  \right)U_{n-i}U_{n-j}
 +\sum_{s\geq 1}(j-i+2s)U_{n-j-s}U_{n-i+s}.   \label{A02}
\end{align}
\begin{rmk}\label{rmk:technique}
In fact, the equalities \eqref{A01}--\eqref{A02} hold true even for all $i,j\in\bbZ_{\geq 0}$
in the sense of the convention \eqref{short notation}.
Details are left to the readers.
\end{rmk}

In order to calculate the central invariants of the bihamiltonian structure $(\mcalP_1, \mcalP_2)$,
we need to find its canonical coordinates $u^i$,
and then compute the corresponding functions
$f^i, H^{ij}_\ell$ and $K^{ii}_\ell$ \eqref{diag fi}--\eqref{Hij and Kij}.
For preparation, let us introduce the symbol $\lmd(p)$ associated to the Lax operator $L$ given in \eqref{n1RR2T-Lax} as follows:
\begin{equation}\label{lmd and fai}
  \lmd(p):=\frac{p}{p-W}\fai(p),\qquad
  \fai(p):=\sum_{i=0}^{n}U_{n-i}p^i,
\end{equation}
where $p\in\bbC$ is a complex parameter. What is more,
for $\ell=1,2$, we introduce the \emph{generating kernel} of the Hamiltonian structure $\mcalP_\ell$ (recall \eqref{general HamP}) as:
\begin{equation}
\label{G-ell-pqe}
  G_\ell(p,q;\veps) := \sum_{k\geq 0}\sfA_{k;\ell}\left(\td\lmd(p), \td\lmd(q)\right)\veps^k,
\end{equation}
where $p,q\in\bbC$ are complex parameters. 


\begin{prop}
The generating kernels $G_\ell(p,q;\veps)$ introduced above have the form:
\begin{align}
  G_1(p,q;\veps)
&=\,
  X(p,q;\veps)
  \frac{
    q\big(\lmd(p\rme^\veps) - \lmd(p)\big)
   -p\rme^\veps\big(\lmd(q)-\lmd(q\rme^{-\veps})\big)
  }
  {\veps\left(p\rme^\veps - q\right)},  \label{G1-pqe}
\\[8pt]
  G_2(p,q;\veps)
&=\,
  X(p,q;\veps)
  \Bigg(
    \frac{
      \big(\lmd(p\rme^\veps) - \lmd(p)\big)
      \big(\lmd(q) - \rme^{n\veps}\lmd(q\rme^{-\veps})\big)
    }{\veps\left(\rme^{n\veps} - 1\right)}
\notag\\[8pt]
&\qquad
  +\frac{p\rme^\veps}{\veps\left(p\rme^\veps - q\right)}
  \big(
    \lmd(p\rme^\veps)\lmd(q\rme^{-\veps})
   -\lmd(p)\lmd(q)
  \big)
  \Bigg), \label{G2-pqe}
\end{align}
where the function $X(p,q;\veps)$ is defined by
\begin{equation*}
  X(p,q;\veps):=
  \frac{\left(p\rme^\veps - W\right)\left(q\rme^{-\veps}-W\right)}{(p-W)(q-W)}.
\end{equation*}
\end{prop}

\begin{proof}
We first consider the generating functions $\sum_{k\geq 0}\sfA_{k;\ell}\left(\td\fai(p), \td\fai(q)\right)\veps^k$ with $\ell=1,2$.
To this end, from the definitions \eqref{general HamP}--\eqref{tensor-Akl} of $\sfA_{k;\ell}$,
it is clear that
\begin{align}
&\,
\sum_{k\geq 0}\sfA_{k;1}(\td U_{n-i}, \td U_{n-j})\veps^k \notag \\
=&\,
  \boldsymbol{1}_{i>0}\boldsymbol{1}_{j>0}U_{n-i-j}\frac{\rme^{i\veps}-\rme^{-j\veps}}{\veps}
 +WU_{n-1-i-j}\frac{\rme^{-(j+1)\veps}-\rme^{(i+1)\veps}}{\veps},
\label{Ak1-ij-gen}\\[12pt]
&\,
\sum_{k\geq 0}\sfA_{k;2}(\td U_{n-i}, \td U_{n-j})\veps^k \notag\\
=&\,
\frac{1}{\veps}
  \left(
    U_{n-i}U_{n-j}
    \frac{\left(\rme^{i\veps}-1\right)\left(1-\rme^{(n-j)\veps}\right)}{\rme^{n\veps}-1}
   +\sum_{s\geq 1}U_{n-i+s}U_{n-j-s}
   \left(
     \rme^{s\veps} - \rme^{(i-j-s)\veps}
   \right)
  \right)\label{Ak2-ij-gen}
\end{align}
hold true for all $0\leq i,j\leq n-1$,
which are obtained by applying the replacement $\Lambda\mapsto\rme^\veps$ to \eqref{260720-1846-1} and \eqref{HamP2Op1}
after the rescaling \eqref{rescaling}.
Indeed, \eqref{Ak1-ij-gen}--\eqref{Ak2-ij-gen} hold true even for all $i,j\in\bbZ_{\geq 0}$
in the sense of the convention \eqref{short notation},
which is analogous to Remark \ref{rmk:technique}.
From the following elementary algebraic identities
\begin{equation*}
\sum_{i,j\geq 0}U_{n-i-j}p^i q^j  =  \frac{p\fai(p)-q\fai(q)}{p-q},\quad
\sum_{i,j\geq 0}U_{n-1-i-j}p^i q^j =  \frac{\fai(p)-\fai(q)}{p-q},
\end{equation*}
\begin{equation*}
  \sum_{i,j\geq 0}\sum_{s\geq 1}
    r^s U_{n-i+s} U_{n-j-s} p^i q^j
=
  \frac{pr}{pr-q}
  \fai(p)
  \Big(\fai(pr)-\fai(q)\Big)
\end{equation*}
for any parameters $p,q$ and $r$, we immediately obtain
\begin{align}
\sum_{k\geq 0}\sfA_{k;1}\left(\td\fai(p), \td\fai(q)\right)\veps^k =&\,
  \frac{1}{\veps}\sum_{i,j\geq 1}
  U_{n-i-j}
  \left(
    \left(p\rme^{\veps}\right)^i q^j
   -p^i\left(q\rme^{-\veps}\right)^j
  \right)\notag \\
&\quad
  +\frac{W}{\veps}\sum_{i,j\geq 0}
    U_{n-1-i-j}
    \left(
      \rme^{-\veps} p^i\left(q \rme^{-\veps}\right)^j
     -\rme^{\veps} \left(p \rme^\veps\right)^i q^j
    \right) \notag \\
=&\, \frac{1}{\veps}
\Bigg(\frac{p\rme^\veps\fai(p\rme^\veps)-q\fai(q)}{p\rme^\veps-q}-\fai(p\rme^\veps)-\fai(q)    \notag\\
&\quad -\frac{p\fai(p)-q\rme^{-\veps}\fai(q\rme^{-\veps})}{p-q\rme^{-\veps}}
  +\fai(p) + \fai(q\rme^{-\veps})
\notag \\
&\quad  +\frac{\rme^{-\veps} W\Big(\fai(p)-\fai(q\rme^{-\veps}) \Big) }{p - q\rme^{-\veps}} - \frac{\rme^{\veps} W\Big(\fai(p\rme^{\veps})-\fai(q) \Big)}{ p\rme^\veps - q } \Bigg)
\notag \\
=&\,
  \frac{
    q \Big(
    \fai\left(p\rme^\veps\right) - \fai(p)
  \Big)
  -p\rme^\veps\Big(
    \fai(q) - \fai(q\rme^{-\veps})
  \Big)
  }
  {\veps\left(p\rme^\veps - q\right)}\notag \\
&\quad
  +\frac{W}{\veps\left(p\rme^\veps - q\right)}
  \Big(
    \fai(p)-\fai(q\rme^{-\veps})
   -\rme^\veps
     \Big(
       \fai(p\rme^\veps) - \fai(q)
     \Big)
  \Big),  \label{F1pqe}
\\[12pt]
\sum_{k\geq 0}\sfA_{k;2}\left(\td\fai(p), \td\fai(q)\right)\veps^k =
&\,
  \frac{1}{\veps\left(\rme^{n\veps}-1\right)}
  \sum_{i,j\geq 0}U_{n-i}U_{n-j}
  \left(
    \rme^{i\veps}-\rme^{(n+i-j)\veps} - 1 + \rme^{(n-j)\veps}
  \right) p^i q^j \notag \\
&\quad
  +\frac{1}{\veps}
    \sum_{i,j\geq 0}\sum_{s\geq 1}
      \rme^{s\veps}U_{n-i+s}U_{n-j-s} p^i q^j  \notag\\
&\quad
  -\frac{1}{\veps}
    \sum_{i,j\geq 0}\sum_{s\geq 1}
      \rme^{-s\veps} U_{n-i+s} U_{n-j-s} \left(p\rme^\veps\right)^i \left(q \rme^{-\veps}\right)^j \notag\\
=&\,
  \frac{
    \Big(
      \fai(p\rme^\veps)-\fai(p)
    \Big)
    \Big(
      \fai(q) - \rme^{n\veps} \fai(q\rme^{-\veps})
    \Big)
  }
  {\veps\left(\rme^{n\veps}-1\right)}  \notag\\
&\quad
  +\frac{p\rme^\veps}{\veps\left(p\rme^{\veps}-q\right)}
  \Big(
    \fai(p\rme^\veps) \fai(q\rme^{-\veps})
   -\fai(p)\fai(q)
  \Big). \label{F2pqe}
\end{align}

Next, from \eqref{HamP1-afa-W} and \eqref{HamP2-afa-W}--\eqref{260720-1846-2}, we have
\begin{align}
\sum_{k\geq 0} \sfA_{k;1}(\td\fai(p), \td W)\veps^k &=\, \frac{\rme^\veps - 1}{\veps} W, \label{260826-1452-1}\\
\sum_{k\geq 0} \sfA_{k;1}(\td W, \td \fai(q))\veps^k &=\, \frac{1-\rme^{-\veps}}{\veps} W, \notag\\
\sum_{k\geq 0} \sfA_{k;1}(\td W, \td W)\veps^k &=\, 0,
\notag\\
\sum_{k\geq 0} \sfA_{k;2}(\td\fai(p), \td W)\veps^k &=\,
  \frac{W\left(\rme^\veps - 1\right)}{\veps\left(\rme^{n\veps}-1\right)}
  \Big(
    \rme^{n\veps}\fai(p) - \fai(p\rme^{\veps})
  \Big), \notag\\
\sum_{k\geq 0} \sfA_{k;2}(\td W, \td \fai(q))\veps^k &=\,
  \frac{W\left(1-\rme^{-\veps} \right)}{\veps\left(\rme^{n\veps}-1\right)}
  \Big(
    \rme^{n\veps}\fai(q\rme^{-\veps}) - \fai(q)
  \Big), \notag\\
\sum_{k\geq 0} \sfA_{k;2}(\td W, \td W)\veps^k &=\,
  W^2\frac{\left(\rme^{n\veps}-\rme^{-\veps}\right)\left(\rme^\veps - 1\right)}
  {\veps\left(\rme^{n\veps}-1\right)}. \label{260826-1452-2}
\end{align}

Finally, it is clear that
\begin{align}
&\,
  G_\ell(p,q;\veps)
=
  pq \, \sum_{k\geq 0}
    \sfA_{k;\ell}\left(
    \td\frac{\fai(p)}{p-W}\,,\,\td\frac{\fai(q)}{q-W}
  \right) \veps^k
\notag\\
=&\,
  \frac{1}{(p-W)(q-W)}
  \sum_{k\geq 0}
  \Big(
    pq\,\sfA_{k;\ell}(\td\fai(p),\td\fai(q))
\notag\\
&\quad
  +\lmd(p)q\,\sfA_{k;\ell}(\td W,\td\fai(q))
  +p\lmd(q)\,\sfA_{k;\ell}(\td\fai(p),\td W)\phantom{\Big(}
\notag\\
&\quad
  +\lmd(p)\lmd(q)\,\sfA_{k;\ell}(\td W,\td W)
\Big) \veps^k.
\end{align}
Substituting \eqref{F1pqe}--\eqref{260826-1452-2}
into this equation,
and using \eqref{lmd and fai} to express $\fai(z)$ in terms of $\lmd(z)$,
we obtain the theorem after a straightforward computation.
\end{proof}

\subsection{Central invariants of $(\mcalP_1, \mcalP_2)$}
\label{subsection:proof of the main theorem}

Recall the symbol $\lambda(p)$ in \eqref{lmd and fai}.
Let $p_i=p_i(\bsv)$, $1\leq i\leq n+1$, be all the roots of the equation $\lmd'(p)=0$,
where $(\,)':=\pp p$,
and assume that they are pairwise distinct and not equal to $0$ or $W$.
Denote the corresponding critical values by
\begin{equation}\label{def of ui}
u^i:=\lmd(p_i),\quad 1\leq i\leq n+1,
\end{equation}
and assume that they are also pairwise distinct. We proceed to show that $\bsu:=(u^1,\dots$, $u^{n+1})$ turns out to be a family of canonical coordinates for the bihamiltonian structure $(\mcalP_1,\mcalP_2)$ given by \eqref{260720-1846-1}--\eqref{260720-1846-2}, and
then find the functions $H^{ij}_{\ell}$ and $K^{ii}_\ell$ in \eqref{Hij and Kij} of $\bsu$, which lead to the central invariants $\mfkc_i$ due to \eqref{def of cInv}.

Firstly, since $\lmd'(p_i)=0$ for $1\leq i\leq n+1$, we note that
\begin{align}
&\,
  \sfA_{k;\ell}\left(\td u^i, \td u^j\right)
=
  \sfA_{k;\ell}\Big(\td\,(\lmd(p_i)) , \td\,(\lmd(p_j))\Big)
\notag\\
=&\,
  \sfA_{k;\ell}\left(\td\lmd(p)\big|_{p\mapsto p_i} + \lmd'(p_i)\td p_i \,,\,
  \td\lmd(q)\big|_{q\mapsto p_j} + \lmd'(p_j)\td p_j\right)
\notag\\
=&\,
  \sfA_{k;\ell}\left(\td\lmd(p),
  \td\lmd(q)\right)\Big|_{(p,q)\mapsto(p_i,p_j)}  \label{Auiuj}
\end{align}
for all $1\leq i,j\leq n+1$, $k\in\bbZ_{\geq 0}$ and $\ell=1,2$.

\begin{prop}\label{prop:can coord}
The bihamiltonian structure $(\mcalP_1,\mcalP_2)$ given by \eqref{260720-1846-1}--\eqref{260720-1846-2} of the $(n,1)$-RR2T
  is semisimple,
  and $\bsu:=(u^1,\dots,u^{n+1})$ introduced in \eqref{def of ui}
  form a family of canonical coordinates.
  Moreover, under such canonical coordinates the leading contravariant metrics $g_\ell$ are given by \eqref{diag fi} with
  \begin{equation} \label{explicit fi}
    f^i = p_i^2\lmd''(p_i),\quad 1\leq i\leq n+1.
  \end{equation}
\end{prop}

\begin{proof}
Comparing the coefficients of $\veps^0$ on both sides of \eqref{G1-pqe}--\eqref{G2-pqe},
and using \eqref{Auiuj} with $k=0$, we arrive at
\begin{equation}\label{g1ij-g2ij}
  g_1(\td u^i,\td u^j)=\delta_{ij}p_i^2\lmd''(p_i),\quad
  g_2(\td u^i,\td u^j)=\delta_{ij}u^ip_i^2\lmd''(p_i)
\end{equation}
for $1\leq i,j\leq n+1$. Thus the proposition is proved.
\end{proof}

Now we arrive at the main theorem of the present section.
\begin{thm}\label{thm:main thm}
The central invariants of the $(n,1)$-type RR2T bihamiltonian structure $(\mcalP_1,\mcalP_2)$
given by \eqref{n1RR2T Pell}--\eqref {260720-1846-2} are all equal to $\frac{1}{24}$.
\end{thm}

\begin{proof}
Comparing the coefficients of $\veps^1$ and $\veps^2$ on both sides of \eqref{G1-pqe}--\eqref{G2-pqe},
and then using \eqref{Hij and Kij} and \eqref{Auiuj}, we arrive at
\begin{align}
  H_1^{ij} &=\,  \frac{p_i p_j}{2(p_i-p_j)}\Big(p_i\lmd''(p_i) + p_j\lmd''(p_j)\Big), \label{H1ij}\\
  H_2^{ij} &=\,
    \frac{p_i p_j}{2(p_i-p_j)}\Big(p_i\lmd''(p_i)u^j + p_j\lmd''(p_j)u^i\Big), \label{H2ij}\\
  K_1^{ii} &=\,  \frac {p_i^2\lmd''(p_i)}{6} + \frac{p_i^3\lmd'''(p_i)}{3} + \frac{p_i^4\lmd^{(4)}(p_i)}{12}
    -\frac{p_i^3\lmd''(p_i)W}{(p_i-W)^2}, \\
  K_2^{ii} &=\,  \frac{n-1}{4n}p_i^4\left(\lmd''(p_i)\right)^2 + u^i K_1^{ii} \label{K2ii}
\end{align}
for each $1\leq i\neq j\leq n+1$.
Following from \eqref{def of cInv} and \eqref{explicit fi}, we have
\begin{align}
  \mfkc_i &=\,  \frac{1}{3\left(f^i\right)^2}
  \left(
    \frac{n-1}{4n}\left(f^i\right)^2
   +\sum_{k\neq i}
     \frac{\left(H_2^{ki} - u^i H_1^{ki}\right)^2}{\left(u^k - u^i\right)f^k}
  \right) \notag \\
&=\,
  \frac{n-1}{12 n} + \frac{1}{12}\sum_{k\neq i}
  \frac{\lmd(p_k)-\lmd(p_i)}{\left(p_k-p_i\right)^2\lmd''(p_k)}
\notag \\
&=\,
  \frac{n-1}{12n} + \frac{1}{12}\left(\frac 1n - \frac 12\right) = \frac{1}{24}, \label{ciRR2D}
\end{align}
where the third equality is obtained by applying the residue theorem to the meromorphic functions on the Riemann sphere:
\[
  m_i(z) := \frac{\lmd(z) - \lmd(p_i)}{(z-p_i)^2\lmd'(z)}.
\]
The theorem is proved.
\end{proof}

This theorem 
generalizes the corresponding results for $n=1,2$ obtained in \cite{AL-triham, 2-1 RR2T} to that for arbitrary $n\geq 1$.

\begin{rmk}
One can also compute the central invariants of the bihamiltonian structure $(\mcalP_2,\mcalP_1)$,
with $\mcalP_2$ regarded as the first Hamiltonian structure and $\mcalP_1$ as the second one.
In this case, a family of canonical coordinates is given by
\begin{equation}\label{utilde}
  \tilde{\bsu}:=(\tilde u^1,\dots,\tilde u^{n+1}),\quad
  \tilde u^i := \frac{1}{u^i} = \frac{1}{\lmd(p_i)}.
\end{equation}
Thanks to \eqref{tensor-Akl}--\eqref{Hij and Kij}, for $\ell=1,2$, we introduce the notations:
\begin{align*}
  \tilde g_\ell^{ij} &:=\, g_{3-\ell}(\td\tilde u^i, \td\tilde u^j)
= \left(\tilde u^i \tilde u^j\right)^2 g_{3-\ell}(\td u^i, \td u^j),
\\
  \tilde H^{ij}_\ell &:=\, H_{3-\ell}(\td\tilde u^i, \td\tilde u^j)
= \left(\tilde u^i \tilde u^j\right)^2 H_{3-\ell}^{ij},
\\
  \tilde K^{ij}_\ell &:=\, K_{3-\ell}(\td\tilde u^i, \td\tilde u^j)
= \left(\tilde u^i \tilde u^j\right)^2 K_{3-\ell}^{ij},
\end{align*}
with $1\leq i,j\leq n+1$.
More exactly, from \eqref{g1ij-g2ij} and \eqref{utilde}, we have
\begin{equation}
\tilde g_1^{ij} = \delta_{ij}\tilde f^i,\quad
\tilde g_2^{ij} = \delta_{ij}\tilde u^i\tilde f^i,\quad \text{where}\quad
\tilde f^i := \left(\tilde u^i\right)^3 p_i^2 \lmd''(p_i).
\end{equation}
Thanks to \eqref{H1ij}--\eqref{H2ij} and \eqref{K2ii}, we obtain
\begin{align*}
  \tilde H_1^{ij}
&=\,
  \frac{p_i p_j \tilde u^i \tilde u^j}{2(p_i-p_j)}
  \Big(
    p_i \tilde u^i \lmd''(p_i)
   +p_j \tilde u^j \lmd''(p_j)
  \Big),
\\
  \tilde H_2^{ij}
&=\,
  \frac{p_i p_j \left(\tilde u^i \tilde u^j\right)^2}{2(p_i - p_j)}
  \Big(
    p_i \lmd''(p_i) + p_j \lmd''(p_j)
  \Big),
\\
  \tilde K_1^{ii}
&=\,
  \frac{n-1}{4n}p_i^4\left(\tilde u^i\right)^4\left(\lmd''(p_i)\right)^2
 +\left(\tilde u^i\right)^{-1}\tilde K_2^{ii}
\end{align*}
for $1\leq i\neq j\leq n+1$.
In the same way as \eqref{ciRR2D}, the central invariants of $(\mcalP_2, \mcalP_1)$ are
\begin{align}
  \tilde \mfkc_i &=\,
  \frac{1}{3(\tilde f^i)^2}
  \left(
    \tilde K_2^{ii} - \tilde u^i \tilde K_1^{ii}
   +\sum_{k\neq i}
     \frac{(\tilde H_2^{ki} - \tilde u^i \tilde H_1^{ki})^2}
     {(\tilde u^k - \tilde u^i) \tilde f^k}
  \right)
\notag \\
&=\,
  -\frac{1}{12\tilde u^i}
  \left(
    \frac{n-1}{n}
   +\sum_{k\neq i}\frac{\lmd(p_k)-\lmd(p_i)}{(p_k - p_i)^2\lmd''(p_k)}
  \right)
=
  -\frac{1}{24\tilde u^i}. \label{central-inv-for-P2P1}
\end{align}
with $1\le i\le n+1$. This result agrees with a general formula in Theorem 5.2 of \cite{Hamiltonian perturbation}.
\end{rmk}

\section{Generating  kernel method for some other lattice hierarchies}


The central invariants of the bihamiltonian structures of the extended bi-graded Toda hierarchies \cite{Carlet-EBTH,CDZ} have been obtained by Carlet, Dubrovin and Zhang \cite{CDZcom}, which are all equal to $\frac{1}{24}$, though no proof is found in print to our best knowledge. This result was quoted in \cite{central inv of cKP}, where Liu, Zhang and Zhou computed the central invariants for constrained KP hierarchies by representing their bihamiltonian structures into multiple integrals.
With a similar method as in \cite{central inv of cKP}, Cao \cite{CZL} showed that the central invariants for the $q$-deformed Gelfand--Dickey hierarchy \cite{rational-reduction, Frenkel-qKdV, Romano} are all equal to $\frac{1}{24}$.

In this section, we want to calculate the central invariants of the bihamiltonian structures of the extended bi-graded Toda hierarchy and of the $q$-deformed Gelfand--Dickey hierarchy, respectively, based the generating kernel of Hamiltonian structure introduced in the previous section. Our method is different from that used in \cite{central inv of cKP}, and it leads to the same results as those in \cite{CZL,CDZcom}. These examples will show that the generating kernel method is efficient at least for the case of lattice integrable hierarchies.



\subsection{Central invariants for extended bi-graded Toda hierarchy}

Given two positive integers $n$ and $m$, the Lax operator $L$ of the extended $(n,m)$ bi-graded Toda hierarchy has the form
\begin{equation}\label{EBTH-Lax}
  L = \Lmd^n + V_{n-1}\Lmd^{n-1} + \cdots + V_{-m}\Lmd^{-m},
\end{equation}
where 
$\Lmd:=\rme^{\veps\p_x}$ is the shift operator as in the previous section,
and $\bsv=(V_{-m}, V_{-m+1}, \dots, V_{n-1})$ is a family of local coordinates on some $(m+n)$-dimensional manifold.
As shown in \cite{Carlet-EBTH} (cf. \cite{BM}),
this hierarchy admits the following compatible brackets
\begin{equation}\label{EBTH-ham-map-form}
  \{F,G\}_\ell := \int\Res\left(\frac{\delta F}{\delta L}
  \left(
    \bfP_{\ell}\frac{\delta G}{\delta L}
  \right)\right)\td x,\quad \ell=1,2
\end{equation}
for arbitrary local functionals $F$ and $G$,
where $\frac{\delta F}{\delta L}:=\sum_{k=-m}^{n-1}\Lmd^{-k}\circ\frac{\delta F}{\delta V_k}$,
and the corresponding Hamiltonian maps $\bfP_{\ell}$ have the form
\begin{align*}
  \bfP_{1}(X) &=\, \left[X_+,L\right]_{\leq 0} - \left[X_-,L\right]_{>0},
\\
  \bfP_{2}(X)
&=\,
  \frac12\left[
    L, (LX+XL)_-
  \right]
 -\frac12 L [L,X]_{\leq 0}
 -\frac12 [L,X]_{\leq 0} L \\
&\quad
 +\frac12\left[L, \frac{\Lmd^n+1}{\Lmd^n-1}\left(\Res[L,X]\right)\right].
\end{align*}
These brackets can be rewritten as
\[
  \{F,G\}_{\ell} = \int \sum_{i,j=-m}^{n-1}\frac{\delta F}{\delta V_i}
  \left(
    \mcalP_\ell^{ij}\frac{\delta G}{\delta V_j}
  \right)\td x,\quad \ell=1,2,
\]
where the entries of the operator-valued matrices
$\mcalP_\ell=\left(\mcalP_\ell^{ij}\right)_{-m\leq i,j\leq n-1}$ 
have the form (recall \eqref{oneS})
\begin{align}
  \mcalP_1^{ij}
&=\,
  \left( \boldsymbol{1}_{i>0}\boldsymbol{1}_{j>0}- \boldsymbol{1}_{i\leq0}\boldsymbol{1}_{j\leq0}\right)\left(V_{i+j} \Lmd^i - \Lmd^{-j}V_{i+j}\right),
\label{EBTH-P1ij} \\
  \mcalP_2^{ij}
&=\,
  V_i\frac{\left(\Lmd^i - \Lmd^n\right)\left(\Lmd^{-j} - 1\right)}{1-\Lmd^n} V_j
+
  \sum_{s<j}
  \left(
    V_{i+j-s}\Lmd^{i-s}V_s - V_s \Lmd^{s-j} V_{i+j-s}
  \right),
\label{EBTH-P2ij}
\end{align}
with $-m\leq i,j\leq n-1$.
Note that throughout this subsection, the following the conventions are used:
\begin{equation}\label{convention-EBTH}
  V_i:=0 ~~ \text{if}~~ i<-m~~\text{or}~~i>n,\quad
  V_n:=1.
\end{equation}

After the rescaling $\mcalP_\ell\mapsto\frac{1}{\veps}\mcalP_\ell$,
the Hamiltonian structures $\mcalP_\ell$ have the standard form \eqref{general HamP}.
Now let us study the central invariants of $(\mcalP_1,\mcalP_2)$ following the approach developed in the previous section.

We start with the symbol
\begin{equation}
  \lmd(p) := p^n + \sum_{i=-m}^{n-1} V_i p^i,\quad p\in\bbC^\times,
\end{equation}
corresponding to the Lax operator \eqref{EBTH-Lax}.
Let $p_1,\dots, p_{n+m}$ be all the roots of $\lmd'(p):=\pfrac{\lmd(p)}{p}$,
and assume that they are non-zero and pairwise distinct.
Similar as before, it turns out that the corresponding critical values
\[
  \bsu:=(u^1,\dots, u^{n+m}),\quad u^i:=\lmd(p_i),
\]
form the canonical coordinates of the bihamiltonian structure $(\mcalP_1,\mcalP_2)$.

From \eqref{EBTH-P1ij}--\eqref{EBTH-P2ij}, we immediately obtain
\begin{align}
  \sum_{k\geq 0}\sfA_{k;1}(\td V_i, \td V_j)\veps^k
&=\,
  \left( \boldsymbol{1}_{i>0}\boldsymbol{1}_{j>0}- \boldsymbol{1}_{i\leq0}\boldsymbol{1}_{j\leq0}\right)V_{i+j}\frac{\rme^{i\veps}-\rme^{-j\veps}}{\veps},
\label{EBTH-pG1}\\
  \sum_{k\geq 0}\sfA_{k;2}(\td V_i, \td V_j)\veps^k
&=\,
  V_i V_j
  \frac{
    \left(\rme^{i\veps} - \rme^{n\veps}\right)
    \left(1-\rme^{-j\veps}\right)
  }
  {\veps\left(\rme^{n\veps}-1\right)}
 +\sum_{s<j}
  V_s V_{i+j-s} \frac{\rme^{(i-s)\veps} - \rme^{(s-j)\veps}}{\veps},
\label{EBTH-pG2}
\end{align}
for $-m\leq i,j\leq n$.
Indeed, it can be verified carefully that
\eqref{EBTH-pG1}--\eqref{EBTH-pG2} hold true even for all $i,j\in\bbZ$
in the sense of \eqref{convention-EBTH}.
Using the following elementary algebraic identities
\begin{align*}
  &\sum_{i,j>0}V_{i+j} p^i q^j
=\,
  \frac{q\lmd_{>0}(p) - p \lmd_{>0}(q)}{p-q},\quad \text{where}\,\,
  \lmd_{>0}(p):=\sum_{i>0}V_i p^i,
\\
  &\sum_{i,j\leq 0}V_{i+j} p^i q^j
=\,
  \frac{p\lmd_{\leq 0}(q) - q \lmd_{\leq 0}(p)}{p-q},\quad \text{where}\,\,
  \lmd_{\leq 0}(p):=\sum_{i \leq 0}V_i p^i,
\\
  &\sum_{i,j\in\bbZ}
  \sum_{s<j}
    r^s V_s V_{i+j-s} p^i q^j
=\,
  \frac{q\lmd(p)\lmd(qr)}{p-q},
\end{align*}
with parameters $p,q$ and $r$, we derive the generating kernels of $\mcalP_\ell$ as
\begin{align}
  G_1(p,q;\veps)
&:=\,
  \sum_{k\geq 0}\sfA_{k;1}(\td\lmd(p), \td\lmd(q))\veps^k
=
  \sum_{i,j\in\bbZ}
    \left(
      \sum_{k\geq 0}
      \sfA_{k;1}(\td V_i, \td V_j)\veps^k
    \right)p^i q^j
\notag \\
&\phantom:=\,
  \frac{
    q\Big(\lmd(p\rme^\veps) - \lmd(p)\Big)
   -p\rme^\veps\Big(\lmd(q) - \lmd(q\rme^{-\veps})\Big)
  }
  {\veps\left(p\rme^\veps - q\right)},
\\
  G_2(p,q;\veps)
&:=\,
  \sum_{k\geq 0}\sfA_{k;2}(\td\lmd(p), \td\lmd(q))\veps^k
=
  \sum_{i,j\in\bbZ}
    \left(
      \sum_{k\geq 0}
      \sfA_{k;2}(\td V_i, \td V_j)\veps^k
    \right)p^i q^j
\notag \\
&\phantom:=\,
  \frac{
    \Big(
      \lmd(p\rme^\veps) - \rme^{n\veps}\lmd(p)
    \Big)
    \Big(
      \lmd(q) - \lmd(q\rme^{-\veps})
    \Big)
  }
  {\veps\left(\rme^{n\veps} - 1\right)}
\notag\\
&\qquad
  + \frac{q}{\veps\left(p\rme^\veps - q\right)}
  \Big(
    \lmd(p\rme^\veps)\lmd(q\rme^{-\veps}) - \lmd(p)\lmd(q)
  \Big).
\end{align}
Comparing the coefficients of $\veps^0$
and then using the relations analogous to \eqref{Auiuj},
we have
\begin{equation}\label{EBTH gij}
  g_1(\td u^i, \td u^j)=\delta_{ij}f^i,\quad
  g_2(\td u^i, \td u^j)=\delta_{ij}u^if^i,\quad
  \text{where}\,\,
  f^i := p_i^2\lmd''(p_i),
\end{equation}
while, the coefficients of $\veps^1$ and $\veps^2$ yield
\begin{align}
  H_1^{ij} &=\,  \frac{p_i p_j}{2(p_i-p_j)}\Big(p_i\lmd''(p_i) + p_j\lmd''(p_j)\Big),  \label{EBTH-H1ij}\\
  H_2^{ij} &=\,
    \frac{p_i p_j}{2(p_i-p_j)}\Big(p_i\lmd''(p_i)u^j + p_j\lmd''(p_j)u^i\Big), \\
  K_1^{ii} &=\,  \frac {p_i^2\lmd''(p_i)}{6} + \frac{p_i^3\lmd'''(p_i)}{3} + \frac{p_i^4\lmd^{(4)}(p_i)}{12}, \\
  K_2^{ii} &=\,  \frac{n-1}{4n}p_i^4\left(\lmd''(p_i)\right)^2 + u^i K_1^{ii} \label{EBTH-K2ij}
\end{align}
for $1\leq i, j\leq n+m$ but $i\neq j$.
Therefore, after a straightforward computation analogous to \eqref{ciRR2D},
we obtain the central invariants 
for the extended $(n,m)$ bi-graded Toda hierarchy as
\[
  \mfkc_1 = \mfkc_2 = \cdots = \mfkc_{n+m} = \frac{1}{24},
\]
which coincide with the results in \cite{CDZcom}.

\subsection{Central invariants for $q$-deformed Gelfand--Dickey hierarchy}

Given an integer $n\ge2$, the Lax operator $L$ of the $q$-deformed Gelfand--Dickey hierarchy has the form
\begin{equation}\label{qGD Lax}
  L := \Lmd^{n} + U_1\Lmd^{n-1} +\cdots + U_{n-1}\Lmd,
\end{equation}
where 
and $\bsv=(U_{1},U_{2},\dots, U_{n-1})$ plays the role of local coordinate on some $n-1$-dimensional manifold.
As shown in \cite{CZL},
the $q$-deformed Gelfand--Dickey hierarchy is equipped with a bihamiltonian structure, given by the compatible Poisson brackets
\begin{equation}\label{EBTH-ham-map-form}
  \{F,G\}_\ell := \int\Res\left(\frac{\delta F}{\delta L}
  \left(
    \bfP_{\ell}\frac{\delta G}{\delta L}
  \right)\right)\td x,\quad \ell=1,2,
\end{equation}
for arbitrary local functionals $F$ and $G$ with
$\frac{\delta F}{\delta L}:=\sum_{k=1}^{n-1}\Lmd^{k-n}\circ\frac{\delta F}{\delta U_k}$. Here the  corresponding Hamiltonian maps $\bfP_{\ell}$ have the form
\begin{align*}
  \bfP_{1}(X) &:=\, \left[L,X\right]_{>0},  \\
  \bfP_{2}(X) &:=\,
    \frac12
      L[L,X]_{>0}
     + \frac12 [L,X]_{>0}L
     - \frac12 \left[L,(LX+XL)_+\right] \notag \\
    &\qquad
     + \frac12 \left[
        \,L\,,\, \frac{\Lmd^{n}+1}{\Lmd^{n}-1}\Res[L,X]
       \,\right].
\end{align*}
Note that the brackets \eqref{EBTH-ham-map-form} can be rewritten as
\[
  \{F,G\}_{\ell} = \int \sum_{i,j=1}^{n-1}\frac{\delta F}{\delta U_i}
  \left(
    \mcalP_\ell^{ij}\frac{\delta G}{\delta U_j}
  \right)\td x,\quad \ell=1,2,
\]
where
\begin{align}
  \mcalP_1^{ij} &:=\, U_{i+j-n}\Lmd^{n-i} - \Lmd^{j-n}U_{i+j-n}, \\
  \mcalP_2^{ij} &:=\,
    U_i\frac{\left(\Lmd^{n-i}-1\right)\left(1-\Lmd^j\right)}{\Lmd^{n}-1}U_j
   +\sum_{s=1}^{j}
    \left(
      U_{j-s}\Lmd^s U_{i+s}
     -U_{i+s}\Lmd^{j-i-s}U_{j-s}
    \right),
\end{align}
with conventions in the present subsection
\begin{equation}\label{convention-qGD}
  U_0:=1,\quad
  U_i:=0~~\text{if}~~ i\ge n.
\end{equation}
More details can be found in \cite{CZL}.

After the rescaling $\mcalP_\ell\mapsto\frac{1}{\veps}\mcalP_\ell$,
the Hamiltonian structures $\mcalP_\ell = \left(\mcalP_\ell^{ij}\right)$
of the $q$-deformed Gelfand--Dickey hierarchy
have the standard form \eqref{general HamP}. We proceed to  compute the central invariants of $(\mcalP_1,\mcalP_2)$ via the generating kernel approach.

We consider the symbol
\begin{equation}
  \lmd(p) := p^{n} + U_1 p^{n-1} + \cdots + U_{n-1} p,\quad p\in\bbC
\end{equation}
corresponding to the Lax operator \eqref{qGD Lax}. Let $p_1,\dots, p_{n-1}$ be all the roots of $\lmd'(p)$, and assume that they are pairwise distinct.
It turns out that the corresponding critical values
$u^i:=\lmd(p_i)$ form the canonical coordinates of the bihamiltonian structure $(\mcalP_1,\mcalP_2)$.

By a computation analogous to those in the previous examples,
we obtain the generating kernels of $\mcalP_\ell$ as
\begin{align*}
  G_1(p,q;\veps)
&=\,
    \frac{
    q \Big(
    \lmd\left(p\rme^\veps\right) - \lmd(p)
  \Big)
  -p\rme^\veps\Big(
    \lmd(q) - \lmd(q\rme^{-\veps})
  \Big)
  }
  {\veps\left(p\rme^\veps - q\right)},
\\[12pt]
  G_2(p,q;\veps)
&=\,
    \frac{
    \Big(
      \lmd(p\rme^\veps)- \lmd(p)
    \Big)
    \Big(
      \lmd(q) - \rme^{ n \veps}  \lmd(q\rme^{-\veps})
    \Big)
  }
  {\veps\left(\rme^{ n \veps}-1\right)}  \notag\\
&\qquad
  +\frac{p\rme^\veps}{\veps\left(p\rme^{\veps}-q\right)}
  \Big(
    \lmd(p\rme^\veps) \lmd(q\rme^{-\veps})
   -\lmd(p)\lmd(q)
  \Big).
\end{align*}
Accordingly, by taking the coefficients of $\veps^0$, $\veps^1$ and $\veps^2$ in $G_\ell(p_i, p_j;\veps)$,
one extracts the required functions $g_\ell^{ij}$, $H_\ell^{ij}$ and $K_\ell^{ij}$ in \eqref{def of cInv},
which have exactly the same expression as \eqref{EBTH gij}--\eqref{EBTH-K2ij} with $\lambda(p)$ replaced by the current version.
Therefore, the central invariants for the
$q$-deformed Gelfand--Dickey hierarchy are
\[
  \mfkc_1 = \mfkc_2 = \cdots = \mfkc_{n-1} = \frac{1}{24}.
\]
This result agrees with that obtained by Cao \cite{CZL} with another method.

\section{Conclusions and remarks}

We have developed a generating kernel method to calculate the central invariants of bihamiltonian structures for lattice hierarchies.
This method is applied to the $(n,1)$-type rational reduction of the 2D Toda hierarchy,
and we have proved that its  central invariants are all equal to $\frac{1}{24}$.
Moreover, this method is tested also valid in the cases of extended bi-graded Toda hierarchy and $q$-deformed Gelfand--Dickey hierarchy,
which leads to the same results as in the literature.
As is shown, the generating kernel method is efficient for calculating the central invariants for lattice hierarchies.

Let us give two remarks. First, it is interesting to ask whether the generating kernel method is valid for differential integrable hierarchies. The answer is positive in some examples, as we have checked, but further research is still needed.
Second, our main result on $(n,1)$-RR2T with $n\ge2$ partially supports the conjecture that  $(n,1)$-RR2T is equivalent to a subhierarchy of the Dubrovin--Zhang hierarchy
associated to a certain $(n+1)$-dimensional generalized Frobenius manifold $\mathcal{M}_{(n,1)}$.
This conjecture has been verified at the dispersionless level.
What is needed is to prove the polynomiality of the structure obtained from the loop equation for $\mathcal{M}_{(n,1)}$,
and to show that its central invariants are all equal to $\frac{1}{24}$.
We will study these topics elsewhere.

\vskip 4ex

\noindent
{\bf Acknowledgments.}
{\noindent \small
The authors thank Professor Youjin Zhang for helpful comments.
The author Qu is partially supported by NSFC No.\,12501323.
The author Wu is partially supported by National Key R\&D Program of China 2023YFA1009801 and
NSFC No.\,12471243.
}

\vskip 4ex

\noindent
{\bf Declaration of AI Use.}
{\noindent \small
No artificial intelligence tool is used to
assist the mathematical research in the main text of this paper.
The Miura transformation in the Appendix is inspired by an AI-generated idea.
}

\renewcommand{\thesection}{A}
\section*{Appendix: Miura-type transformation to the bihamiltonian structure for $(n,1)$-RR2T}
\addcontentsline{toc}{section}{Appendix: Miura-type transformation to the bihamiltonian structure for $(n,1)$-RR2T}

For the $(n,1)$-type rational reduction of the 2D Toda hierarchy, its Lax operator \eqref{n1RR2T-Lax} of reads
\begin{equation}\label{n1RR2T-Lax2}
L=\,
  \left(1-W\Lmd^{-1}\right)^{-1}
  \left(
    \Lmd^n + \sum_{k=1}^{n} U_{k}\Lmd^{n-k}
  \right)
=
  \Lmd^n +
    \sum_{k=1}^{\infty}\widetilde U_k \Lmd^{n-k},
\end{equation}
where $\widetilde U_k$ with $k\ge n+1$ are differential polynomials in $\{ \widetilde U_i \}_{i=1}^{n+1}$. This Lax operator can be regarded as a certain deformation of the one \eqref{EBTH-Lax} for the $(n,1)$-type extended bi-graded Toda hierarchy ($(n,1)$-EBGT for short). Based on this observation, we will find a certain Miura-type transformation between the bihamiltonian structure for $(n,1)$-RR2T and a mild modification of that for $(n,1)$-EBGT.

\subsection{Bihamiltonian structure for $(n,1)$-RR2T in another coordinate}

We want to apply a Miura-type transformation to the bihamiltonian structure $(\mcalP_1, \mcalP_2)$ given by \eqref{260720-1846-1}--\eqref{260720-1846-2} for $(n,1)$-RR2T. Namely, we want represent  $(\mcalP_1, \mcalP_2)$ in an alternate coordinate.

By rewriting \eqref{n1RR2T-Lax2} to the form
\[
  \left(1-W\Lmd^{-1}\right)
  \left(
    \Lmd^n + \sum_{i=1}^{\infty}\widetilde U_i \Lmd^{n-i}
  \right)
=
  \Lmd^n + \sum_{i=1}^{n} U_{i}\Lmd^{n-i},
\]
we have
\begin{align}\label{Miura-transf}
  U_i &=\, \widetilde U_i - W\widetilde U_{i-1}^{[-1]},\quad \forall\, i\geq 1.
\end{align}
For instance, we have
\begin{equation}\label{261105}
W=\frac{\widetilde U_{n+1}}{\widetilde U_n^{[-1]}},\quad
  \widetilde U_{n+1+k} = \widetilde U_{n+1}^{[-k]}\prod_{i=0}^{k-1}W^{[-i]} ~~\text{for}~~ k\geq 1.
\end{equation}
Here and below we still use the conventions \eqref{short notation}.
What is more, we introduce the following additional notations
\begin{equation}\label{short notation - tilde U}
  \widetilde U_0:=1,\quad \widetilde U_k:=0\quad\text{if}\,\, k<0.
\end{equation}

It is easy to see that
$\tilde \bsv := (\widetilde U_1,\dots, \widetilde U_n; \widetilde U_{n+1})$
can be expressed explicitly in terms of $\bsv= (U_1,\dots, U_n; W)$, and vice versa.
That is to say, we have a Miura-type transformation $\bsv\mapsto\tilde\bsv$ yielded by \eqref{Miura-transf}.

Suppose that the bihamiltonian structure $(\mcalP_1,\mcalP_2)$ leads to the following Poisson brackets
under the coordinates $\tilde\bsv$:
\begin{align}\label{261025-1643}
\{F,G\}_\ell =
  \int \sum_{i,j=1}^{n+1}
    \frac{\delta F}{\delta \widetilde U_i}
    \left(
      \widetilde\mcalP_\ell^{ij}\frac{\delta G}{\delta \widetilde U_j}
    \right)\td x,\quad \forall\, F,G\in\mcalF,\, \ell=1,2.
\end{align}
Clearly, the operator-valued matrices $\widetilde\mcalP_\ell=\left(\widetilde\mcalP_\ell^{ij}\right)_{1\leq i,j\leq n+1}$ and  $\mcalP_\ell=\left(\mcalP_\ell^{ij}\right)_{1\leq i,j\leq n+1}$ are related by
\begin{equation}\label{JPJdag}
\widetilde\mcalP_\ell =
\mcalJ\mcalP_\ell\mcalJ^\dag,\quad \ell=1,2,
\end{equation}
where the entries of the Jacobian
$
    \mcalJ =\left(\mcalJ^i_j\right)_{1\leq i,j\leq n+1}
      \in\mcalA\fps{\veps\p_x}\otimes\bbC^{(n+1)\times (n+1)}
$
has the form
\begin{equation}\label{jacobian}
    \mcalJ^i_j := \sum_{s\geq 0}\pfrac{\tilde v^i}{v^{j,s}}\p_x^s,\quad v^{j,s}:=\p_x^s v^j,
\end{equation}
and the entries of $\mcalJ^\dag$ read
\[
\left(\mcalJ^\dag\right)^i_j:=(\mcalJ^j_i)^\dag=\sum_{s\geq 0}(-\p_x)^s\pfrac{\tilde v^j}{v^{i,s}}.
\]
We have the following proposition that will be proved in next subsection.
\begin{prop} \label{thm:tildePell}
Given a positive integer $n$, after the Miura-type transformation given by \eqref{Miura-transf} with $1\le i\le n+1$, the bihamiltonian structure  $(\mcalP_1,\mcalP_2)$ in \eqref{260720-1846-1}--\eqref{260720-1846-2} is converted to $(\widetilde\mcalP_1, \widetilde\mcalP_2)$ with entries
\begin{align}
  \widetilde\mcalP_1^{ij} &:=\,
    (\boldsymbol 1_{i<n}\boldsymbol 1_{j<n}-\boldsymbol 1_{i\geq n}\boldsymbol 1_{j\geq n})
    \left(
      \widetilde U_{i+j-n}\Lmd^{n-i}-\Lmd^{j-n}\widetilde U_{i+j-n}
    \right), \label{tildeP1ij}
\\
  \widetilde\mcalP_2^{ij} &:=\,
    \widetilde U_i
    \frac{\left(\Lmd^{n-i}-\Lmd^n\right)\left(\Lmd^{j-n}-1\right)}{1-\Lmd^n}\widetilde U_j
   +
    \sum_{s\geq 0}
    \left(
      \widetilde U_{j-s}\Lmd^s\widetilde U_{i+s}
     -\widetilde U_{i+s}\Lmd^{j-i-s}\widetilde U_{j-s}
    \right), \label{tildeP2ij}
\end{align}
for $1\leq i,j\leq n+1$.

\end{prop}

As shown in \eqref{261105}, the function $\widetilde U_i$ is non-zero for all $i>n+1$.
If we reset $\widetilde U_i:=0$ for $i>n+1$ in \eqref{tildeP1ij}--\eqref{tildeP2ij},
then the resulting $(\widetilde\mcalP_1,\widetilde\mcalP_2)$ coincides with
the bihamiltonian structure
\eqref{EBTH-P1ij}--\eqref{EBTH-P2ij} for $m=1$,
up to the replacement of coordinates:
\[\widetilde U_i\mapsto V_{n-i},\quad 1\leq i\leq n+1. \]
Therefore, the bihamiltonian structure
$(\widetilde\mcalP_1,\widetilde\mcalP_2)$ introduced by \eqref{tildeP1ij}--\eqref{tildeP2ij} and \eqref{261105}
can be regarded as a modification of that of the $(n,1)$-type extended bi-graded Toda hierarchy.

\begin{ex}
  When $n=1$, the $(1,1)$-RR2T is known as the Ablowitz--Ladik hierarchy, and its bihamiltonian structure $(\mcalP_1,\mcalP_2)$ in \eqref{260720-1846-1}--\eqref{260720-1846-2}
  reads (see \cite{AL-triham,ZQL-n1RR2T} and references therein)
  \[
    \mcalP_1 = \begin{pmatrix}
                 W\Lmd^{-1}-\Lmd W & (\Lmd-1)W \\[9pt]
                 W(1-\Lmd^{-1}) & 0
               \end{pmatrix},\quad
    \mcalP_2 = \begin{pmatrix}
                 0 & U_1(\Lmd-1)W \\[9pt]
                 W(1-\Lmd^{-1})U_1 & W(\Lmd-\Lmd^{-1})W
               \end{pmatrix}.
  \]
After the Miura-type transformation
\[
  \begin{cases}
    \widetilde U_1 &=\, U_1 + W, \\
    \widetilde U_2 &=\, WW^{[-1]} + WU_1^{[-1]}
  \end{cases}
\]
with jacobian
\[
  \mcalJ = \begin{pmatrix}
             1 & 1 \\
             W\Lmd^{-1} &  U_1^{[-1]} + W^{[-1]} + W\Lmd^{-1}
           \end{pmatrix},
\]
the above bihamiltonian structure is converted to $(\widetilde\mcalP_1, \widetilde\mcalP_2)$ in \eqref{tildeP1ij}--\eqref{tildeP2ij}
of the form
  \begin{align}
    \widetilde\mcalP_1 &=\,
      \begin{pmatrix}
        0 & (\Lmd-1)\widetilde U_2 \\[9pt]
        \widetilde U_2(1-\Lmd^{-1}) & \Lmd\widetilde U_3 - \widetilde U_3\Lmd^{-1}
      \end{pmatrix},  \label{ex-P1}\\[12pt]
    \widetilde\mcalP_2 &=\,
      \begin{pmatrix}
        \Lmd\widetilde U_2 - \widetilde U_2\Lmd^{-1}
      & \widetilde U_1(\Lmd-1)\widetilde U_2 + \Lmd^2\widetilde U_3 - \widetilde U_3\Lmd^{-1} \\[9pt]
        \widetilde U_2(1-\Lmd^{-1})\widetilde U_1 + \Lmd\widetilde U_3 - \widetilde U_3\Lmd^{-2}
      &  \Omega
      \end{pmatrix}, \label{ex-P2}
  \end{align}
where
\begin{align}
\Omega &=\,
     \widetilde U_2(\Lmd-\Lmd^{-1})\widetilde U_2
    +\widetilde U_1\Lmd\widetilde U_3 - \widetilde U_3\Lmd^{-1}\widetilde U_1
    +\Lmd^2\widetilde U_4 - \widetilde U_4\Lmd^{-2}, \label{exOmg} \\
  \widetilde U_3 &=
  \frac{\widetilde U_2 \widetilde U_2^{[-1]}}{\widetilde U_1^{[-1]}},
\quad
  \widetilde U_4 =
  \frac{\widetilde U_2 \widetilde U_2^{[-1]}\widetilde U_2^{[-2]}}{\widetilde U_1^{[-1]}\widetilde U_1^{[-2]}}.
\end{align}
Suppose that $\widetilde U_3 = \widetilde U_4=0$ in \eqref{ex-P1}--\eqref{exOmg},
then the resulting $(\widetilde\mcalP_1,\widetilde\mcalP_2)$ coincides with the well-known bihamiltonian structure for the Toda hierarchy.

\end{ex}

\subsection{Proof of Proposition \ref{thm:tildePell}}

 We proceed to prove Proposition \ref{thm:tildePell}.
Firstly, by taking variations on both sides of \eqref{Miura-transf}, we obtain
\[
  \mcalK
  \begin{pmatrix}
    \delta\widetilde U_1 \\
    \vdots \\
    \delta\widetilde U_n \\
    \delta\widetilde U_{n+1}
  \end{pmatrix}
=
  \widetilde\mcalK
  \begin{pmatrix}
    \delta U_1 \\
    \vdots \\
    \delta U_n \\
    \delta W
  \end{pmatrix},
\]
where the operator-valued matrices
\[
  \mcalK
:=
  \begin{pmatrix}
    1 &   &   &   &  \\
    -W\Lmd^{-1} & 1 &   &   &   \\
      & \ddots & \ddots&  &  \\
      &   & -W\Lmd^{-1} & 1 &  \\
      &   &   &   -W\Lmd^{-1}& 1
  \end{pmatrix},
\quad
  \widetilde\mcalK
:=
  \begin{pmatrix}
    1 &   &   &   &  1\\
      & 1 &   &   &  \widetilde U_1^{[-1]} \\
      &   & \ddots&  &  \vdots\\
      &   &   & 1 &  \widetilde U_{n-1}^{[-1]}\\
      &   &   &   &  \widetilde U_n^{[-1]}
  \end{pmatrix}.
\]
In other words, the entries of $\mcalK=\left(\mcalK^i_j\right)$ and
$\widetilde\mcalK = \left(\widetilde\mcalK^i_j\right)$ are
\begin{align}\label{Kij}
  \mcalK^i_j =\,  \delta_{ij} - \delta_{i,j+1}W\Lmd^{-1}, \quad
  \widetilde\mcalK^i_j =\, \delta_{ij}\boldsymbol 1_{j\leq n} + \widetilde U_{i-1}^{[-1]}\delta_{j,n+1}
\end{align}
for $1\leq i,j\leq n+1$. It follows that the Jacobian in \eqref{jacobian} is $\mcalJ=\mcalK^{-1}\widetilde\mcalK$.
Thanks to \eqref{JPJdag}, it suffices to verify
\begin{equation}\label{MiuraRelation}
  \widetilde\mcalK\mcalP_\ell\widetilde\mcalK^\dag
 =\mcalK\widetilde\mcalP_\ell\mcalK^\dag,  \quad \ell=1,2.
\end{equation}

When $\ell=1$, for $1\leq i,j\leq n$, by using \eqref{Kij} and \eqref{260720-1846-1}--\eqref{HamP1-afa-W} we have
\begin{align}
  \left(\widetilde\mcalK\mcalP_1\widetilde\mcalK^\dag\right)^{ij}
&=\,
  \left(
    U_{i+j-n}\Lmd^{n-i}-\Lmd^{j-n}U_{i+j-n}
  \right)\boldsymbol 1_{i<n}\boldsymbol 1_{j<n} \notag
\\
&\quad +\left(
    W\Lmd^{j-n-1}U_{i+j-n-1}-U_{i+j-n-1}\Lmd^{n+1-i}W
  \right)\notag \\
&\quad +\delta_{i,n}(\Lmd-1)W\widetilde U_{j-1}^{[-1]}
+\delta_{j,n}\widetilde U_{i-1}^{[-1]}W(1-\Lmd^{-1}),
\end{align}
while by using \eqref{Miura-transf}, \eqref{tildeP1ij} and \eqref{Kij} we have
\begin{align*}
\left(\mcalK\widetilde\mcalP_1\mcalK^\dag\right)^{ij}
&=\,
  (\boldsymbol 1_{i<n}\boldsymbol 1_{j<n}-\boldsymbol 1_{i\geq n}\boldsymbol 1_{j\geq n})
  \left(
  \widetilde U_{i+j-n}\Lmd^{n-i}-\Lmd^{j-n}\widetilde U_{i+j-n}
  \right)\\
&\quad
  -\boldsymbol 1_{i<n}
  \left(
    \widetilde U_{i+j-n-1}\Lmd^{n-i}-\Lmd^{j-n-1}\widetilde U_{i+j-n-1}
  \right)\Lmd W\\
&\quad
  -\boldsymbol 1_{j<n} W\Lmd^{-1}
  \left(
    \widetilde U_{i+j-n-1}\Lmd^{n+1-i}
   -\Lmd^{j-n}\widetilde U_{i+j-n-1}
  \right)\\
&\quad
  + W\Lmd^{-1}
  \left(
    \widetilde U_{i+j-n-2}\Lmd^{n+1-i}
   -\Lmd^{j-1-n}\widetilde U_{i+j-n-2}
  \right)\Lmd W
\\
=&\,
  \left(
    (\boldsymbol 1_{i<n}\boldsymbol 1_{j<n}-\boldsymbol 1_{i\geq n}\boldsymbol 1_{j\geq n})\widetilde U_{i+j-n}
   -\boldsymbol 1_{j<n}
     W\widetilde U_{i+j-n-1}^{[-1]}
  \right)\Lmd^{n-i}\\
&\quad
  -\Lmd^{j-n}
  \left(
    (\boldsymbol 1_{i<n}\boldsymbol 1_{j<n}-\boldsymbol 1_{i\geq n}\boldsymbol 1_{j\geq n})\widetilde U_{i+j-n}
   -\boldsymbol 1_{i<n}W\widetilde U_{i+j-n-1}^{[-1]}
  \right)\\
&\quad
  -\left(
    \boldsymbol 1_{i<n}\widetilde U_{i+j-n-1} - W\widetilde U_{i+j-n-2}^{[-1]}
  \right)\Lmd^{n+1-i}W\\
&\quad
  +W\Lmd^{j-n-1}
  \left(
    \boldsymbol 1_{j<n}\widetilde U_{i+j-n-1}
   -W\widetilde U_{i+j-n-2}^{[-1]}
  \right)
\\
=&\,
  -\delta_{i,n}\widetilde U_{i+j-n}\Lmd^{n-i}
  +\boldsymbol 1_{j<n} U_{i+j-n}\Lmd^{n-i}\\
&\quad
  +\delta_{j,n}\Lmd^{j-n}\widetilde U_{i+j-n}
  -\boldsymbol 1_{i<n}\Lmd^{j-n}U_{i+j-n}\\
&\quad
  -\left(
    U_{i+j-n-1}-\delta_{i,n}\widetilde U_{i+j-n-1}
  \right)\Lmd^{n+1-i}W\\
&\quad
  +W\Lmd^{j-n-1}\left(
    U_{i+j-n-1}-\delta_{j,n}\widetilde U_{i+j-n-1}
  \right).
\end{align*}
Hence we obtain
\begin{equation}\label{}
  \left(\widetilde\mcalK\mcalP_1\widetilde\mcalK^\dag\right)^{ij}
 = \left(\mcalK\widetilde\mcalP_1\mcalK^\dag\right)^{ij}, \quad 1\leq i,j\leq n.
\end{equation}
The cases of $i=n+1$ or $j=n+1$ can be verified in a similar way,
and we omit the details here.
Thus the equality \eqref{MiuraRelation} holds true for $\ell=1$.

When $\ell=2$, let us write
\[
  \mcalP_2 = \mcalP_{2,1}+\mcalP_{2,2},\quad
  \widetilde\mcalP_{2} = \widetilde\mcalP_{2,1} + \widetilde\mcalP_{2,2},
\]
where $\mcalP_{2,1}$ and $\widetilde\mcalP_{2,1}$
are the ``fractional'' parts, while $\mcalP_{2,2}$ and $\widetilde\mcalP_{2,2}$ are the ``polynomial'' parts. More exactly,   such matrices have entries as follows, for $1\leq i,j\leq n+1$,
\begin{align*}
  \mcalP_{2,1}^{ij}&:=\,
  \begin{cases}
    U_i\frac{\left(\Lmd^{n-i}-1\right)(1-\Lmd^j)}{\Lmd^n-1}U_j, & \quad \text{if} ~~ i,j<n+1; \\[5pt]
    U_i\frac{\left(\Lmd^n-\Lmd^{n-i}\right)(\Lmd-1)}{\Lmd^n-1}W, & \quad \text{if} ~~  i<n+1,\, j=n+1;\\[5pt]
    W\frac{(\Lmd^j-1)(1-\Lmd^{-1})}{\Lmd^n-1}U_j, & \quad \text{if} ~~  i=n+1,\,j<n+1; \\[5pt]
    W\frac{(\Lmd^n-\Lmd^{-1})(\Lmd-1)}{\Lmd^n-1}W, & \quad \text{if} ~~  i,j = n+1,
  \end{cases}
\\
  \widetilde\mcalP_{2,1}^{ij}&:=\,
    \widetilde U_i
    \frac{\left(\Lmd^{n-i}-\Lmd^n\right)\left(\Lmd^{j-n}-1\right)}{1-\Lmd^n}
    \widetilde U_j,
\\
  \mcalP_{2,2}^{ij}&=\,
    \boldsymbol 1_{i\leq n}\boldsymbol 1_{j\leq n}
    \sum_{s\geq 1}
      \left(
        U_{j-s}\Lmd^s U_{i+s}
       -U_{i+s}\Lmd^{j-i-s}U_{j-s}
      \right),
\\
  \widetilde\mcalP_{2,2}^{ij}&=\,
    \sum_{s\geq 0}
    \left(
      \widetilde U_{j-s}\Lmd^s\widetilde U_{i+s}
     -\widetilde U_{i+s}\Lmd^{j-i-s}\widetilde U_{j-s}
    \right).
\end{align*}
In case $1\leq i,j\leq n$,
a straightforward computation leads to
\begin{align*}
  \left(
    \widetilde\mcalK\mcalP_{2,1}\widetilde\mcalK^\dag
  \right)^{ij}
&=\,
  \mcalP_{2,1}^{ij} + U_i\frac{(\Lmd^n-\Lmd^{n-i})(\Lmd-1)}{\Lmd^n-1}W\widetilde U_{j-1}^{[-1]}\\
&\quad
   +W\widetilde U_{i-1}^{[-1]}
   \frac{(\Lmd^j-1)(1-\Lmd^{-1})}{\Lmd^n-1}U_j
   +W\widetilde U_{i-1}^{[-1]}
     \frac{(\Lmd^n-\Lmd^{-1})(\Lmd-1)}{\Lmd^n-1}
     W\widetilde U_{j-1}^{[-1]}\\
&=\,
  U_i
    \frac{\left(\Lmd^{n+1-i}-1\right)(\Lmd^{j-1}-1)}{1-\Lmd^n}
  U_j
+
  U_i
    \frac{\left(\Lmd^{n-i}-\Lmd^{-1}\right)(\Lmd-1)}{1-\Lmd^n}
  \widetilde U_j\\
&\quad +
  \widetilde U_i
    \frac{\left(1-\Lmd^{-1}\right)(\Lmd^{n+1}-\Lmd^j)}{1-\Lmd^n}
  U_j
+
  \widetilde U_i
    \frac{\left(\Lmd^{-1}-1\right)(\Lmd^{n+1}-1)}{1-\Lmd^n}
  \widetilde U_j,
\end{align*}
where the last equality follows from substituting the expression given in \eqref{Miura-transf}.
With the same method, one can compute
$\left(\mcalK\widetilde\mcalP_{2,1}\mcalK^\dag\right)^{ij}$. Hence we obtain
\begin{equation}
\begin{split}
&\,
  \left(\widetilde\mcalK\mcalP_{2,1}\widetilde\mcalK^\dag\right)^{ij}
 -\left(\mcalK\widetilde\mcalP_{2,1}\mcalK^\dag\right)^{ij} \\
=&\,
  U_i(1-\Lmd^{j-i})U_j
 +\left(\widetilde U_i - U_i\right)(1-\Lmd^{-1})\widetilde U_j \\
&\quad
  +\delta_{j,1}
  \Big(
    \widetilde U_i(\Lmd - 1)
    + U_i (1-\Lmd^{1-i})
  \Big)W,
 \quad 1\leq i,j\leq n.
\end{split}\label{tildeP22-1}
\end{equation}
On the other hand, for $1\leq i,j\leq n$, we have
\begin{equation}
\begin{split}
&\,
  \left(\widetilde\mcalK\mcalP_{2,2}\widetilde\mcalK^\dag\right)^{ij}=\mcalP_{2,2}^{ij}\\
=&\,
  \sum_{s\geq 0}
  \left(
    U_{j-s}\Lmd^s U_{i+s}-U_{i+s}\Lmd^{j-i-s}U_{j-s}
  \right)
+ U_i\left(\Lmd^{j-i}-1\right)U_j,
\end{split}\label{tildeP22-2}
\end{equation}
and
\begin{align}
&\,
  \left(\mcalK\widetilde\mcalP_{2,2}\mcalK^\dag\right)^{ij} \notag\\
=&\,
\widetilde P_{2,2}^{ij}
-\boldsymbol 1_{j>1}\widetilde\mcalP_{2,2}^{i,j-1}\Lmd W
-\boldsymbol 1_{i>1} W\Lmd^{-1}\widetilde\mcalP_{2,2}^{i-1,j}
+\boldsymbol 1_{i>1} \boldsymbol 1_{j>1}
 W\Lmd^{-1}\widetilde\mcalP_{2,2}^{i-1,j-1}\Lmd W \notag\\
=&\,
 \left(
   \widetilde\mcalP_{2,2}^{ij} - \boldsymbol 1_{j>1}\widetilde\mcalP_{2,2}^{i,j-1}\Lmd W
 \right)
-W\Lmd^{-1}
 \left(
   \widetilde\mcalP_{2,2}^{i-1,j} - \boldsymbol 1_{j>1}\widetilde\mcalP_{2,2}^{i-1,j-1}\Lmd W
 \right) \notag\\
=&\,
  \sum_{s\geq 0}
  \left(
    \widetilde U_{j-s}\Lmd^s U_{i+s}
   -\widetilde U_{i+s}\Lmd^{j-i-s} U_{j-s}
  \right)
 + \widetilde U_j W \widetilde U_{i-1}^{[-1]}
 + \delta_{j,1}\widetilde U_i(1-\Lmd^{-i})\Lmd W \notag\\
&\quad -
  \sum_{s\geq 0}\left(
      W\widetilde U_{j-s-1}^{[-1]}\Lmd^s U_{i+s}
     -W\widetilde U_{i+s-1}^{[-1]}\Lmd^{j-i-s}U_{j-s} \right)\notag\\
&\quad
  -W\widetilde U_j^{[-1]}\widetilde U_{i-1}^{[-1]}\Lmd^{-1}
  -\delta_{j,1}W\widetilde U_{i-1}^{[-1]}(1-\Lmd^{1-i})W\notag\\
=&\,
  \sum_{s\geq 0}
  \left(
    U_{j-s}\Lmd^s U_{i+s}
   -U_{i+s}\Lmd^{j-i-s} U_{j-s}
  \right)  \notag\\
&\quad
+\left(U_i - \widetilde U_i\right)\left(\Lmd^{-1}-1\right)\widetilde U_j
+
 \delta_{j,1}
\Big(
  \widetilde U_i(\Lmd-1)  + U_i(1-\Lmd^{1-i})
\Big) W,
\label{tildeP22-3}
\end{align}
where the second equality follows from the following identities
\[
    \boldsymbol 1_{i>1}\widetilde\mcalP_{2,2}^{i-1,j} = \widetilde\mcalP_{2,2}^{i-1,j},\quad   1\leq i,j\leq n,
\]
Thus, from \eqref{tildeP22-1}--\eqref{tildeP22-3} it follows that
\[
\left(\widetilde\mcalK\mcalP_{2}\widetilde\mcalK^\dag\right)^{ij}
=\left(\mcalK\widetilde\mcalP_{2}\mcalK^\dag\right)^{ij}, \quad   1\leq i,j\leq n.
\]
The cases of $i=n+1$ or $j=n+1$ can be verified similarly, hence \eqref{MiuraRelation} holds true for $\ell=2$.
Therefore Proposition \ref{thm:tildePell} is proved.

\end{document}